# An Event Horizon Imager (EHI) Mission Concept Utilizing Medium Earth Orbit Sub-mm Interferometry[*]

KUDRIASHOV V[1]    MARTIN-NEIRA M[2]    ROELOFS F[1]    FALCKE H[1,3]
BRINKERINK C[1]    BARYSHEV A[4]    HOGERHEIJDE M[5,6]    YOUNG A[1]
POURSHAGHAGHI H[1]    KLEIN-WOLT M[1]    MOSCIBRODZKA M[1]    DAVELAAR J[1,7]
BARAT I[2]    DUESMANN B[2]    VALENTA V[2]    PERDIGUES ARMENGOL J M[2]
DE WILDE D[2]    MARTIN IGLESIAS P[2]    ALAGHA N[2]    VAN DER VORST M[2]

1(*Department of Astrophysics/IMAPP, Radboud University Nijmegen, P.O. Box 9010, 6500 GL Nijmegen, the Netherlands*)

2(*ESTEC/ESA, Keplerlaan 1, 2201 AZ Noordwijk, the Netherlands*)

3(*Max Planck Institute for Radio Astronomy, Auf dem Hügel 69, D-53121 Bonn (Endenich), Germany*)

4(*University of Groningen, Landleven 12, 9747 AD Groningen, the Netherlands*)

5(*Leiden Observatory, Leiden University, PO Box 9513, 2300 RA, Leiden, the Netherlands*)

6(*Anton Pannekoek Institute for Astronomy, University of Amsterdam, Science Park 904, 1098 XH, Amsterdam, the Netherlands*)

7(*Center for Computational Astrophysics, Flatiron Institute, 162 Fifth Avenue, New York, NY 10010, USA*)

**Abstract**    Submillimeter interferometry has the potential to image supermassive black holes on event horizon scales, providing tests of the theory of general relativity and increasing our understanding of black hole accretion processes. The Event Horizon Telescope (EHT) performs these observations from the ground, and its main imaging targets are Sagittarius A* in the Galactic Center and the black hole at the center of the M87 galaxy. However, the EHT is fundamentally limited in its performance by atmospheric effects and sparse terrestrial $(u, v)$-coverage (Fourier sampling of the image). The scientific interest in quantitative studies of the horizon size and shape of these black holes has motivated studies into using space interferometry which is free of these limitations. Angular resolution considerations and interstellar scattering effects push the desired observing frequency to bands above 500 GHz.

This paper presents the requirements for meeting these science goals, describes the concept of interferometry from Polar or Equatorial Medium Earth Orbits (PECMEO) which we dub the Event Horizon Imager (EHI), and utilizes suitable space technology heritage. In this concept, two or three satellites orbit at slightly different orbital radii, resulting in a dense and uniform spiral-shaped $(u, v)$-coverage over time. The local oscillator signals are shared via an inter-satellite link, and the data streams are correlated on-board before final processing on the ground. Inter-satellite metrology and satellite positioning are extensively employed to facilitate the knowledge of the instrument position vector, and its time derivative. The European space heritage usable for both the front ends and the antenna technology of such an instrument is investigated. Current and future sensors for the required inter-satellite metrology are listed. Intended performance estimates and simulation results are given.

---

\* The research work reported in the paper was partly supported by the Project NPI-552 "Space-to-space Interferometer System to Image the Event Horizon of the Super Massive Black Hole in the Center of Our Galaxy" co-funded by the European Space Agency (ESA) and the Radboud University of Nijmegen (ESA contract 4000122812), and by the NWO project PIPP "Breakthrough Technologies for Interferometry in Space".
  Received August 20, 2020. Revised January 29, 2021
  E-mail: V.Kudriashov@astro.ru.nl. Corresponding author: Kudriashov V

**Classified index** V 476.9, TP 73

# 1 Introduction

Studying the size and shape of a black hole "shadow"[1] is considered to be one of the crucial experimental tests of General Relativity[2-4]. The Supermassive Black Hole (SMBH) in the center of our Galaxy, with the associated radio source Sgr A*, has a mass $M$ of $3.964 \pm 0.047_{\text{stat}} \pm 0.0264_{\text{sys}} \times 10^6 M_\odot$ and is located at a distance $D$ of $7946 \pm 50_{\text{stat}} \pm 32_{\text{sys}}$ pc from Earth[5]. The GRAVITY Collaboration[6] measure a distance of $8178 \pm 13_{\text{stat}} \pm 22_{\text{sys}}$ pc. Assuming a Schwarzschild (non-spinning) black hole, the expected shadow size is $2\sqrt{27}GM/c^2D = 51\,\mu\text{as}$, where $G$ is Newton's gravitational constant and $c$ is the speed of light. This expected angular size is the largest of all black holes in the sky[7]. The shadow of the black hole in the nucleus of the active radio galaxy Messier 87 (M87*) has been imaged with the Event Horizon Telescope (EHT) as an asymmetric ring[8,9]. Model fitting led to an angular size of one gravitational radius $GM/c^2D$ of $3.8 \pm 0.4\,\mu\text{as}$[10,11], leading to a Schwarzschild shadow size of $39 \pm 4\,\mu\text{as}$. The effect of the black hole spin on the angular size of the shadow is $\lesssim 4\%$[12-14]. In addition to fundamental physics, such observations promise a rich harvest for astrophysical studies.

The angular resolution required for decisive observations of the SMBH shadow in these objects is thus at the level of $20\,\mu\text{as}$ or sharper. This resolution can be achieved only by the technique of Very Long Baseline Interferometry (VLBI).

For the longest baselines available on the surface of Earth, the observations reaching the required angular resolution must be conducted at short millimeter and sub-millimeter wavelengths. The results of observing M87* at 230 GHz (1.3 mm) with the EHT have been published recently[8-11,15,16]. The EHT has demonstrated an angular resolution of about $20\,\mu\text{as}$ in the image of M87*. It is expected that, after ongoing expansions and upgrades, the EHT will also operate at 345 GHz (0.87 mm). EHT images are limited in angular resolution by achievable baseline lengths (ultimately – the Earth's diameter) and are affected significantly by propagation effects in the Earth's atmosphere (*e.g.* Ref. [17]). Signal attenuation and phase turbulence introduced primarily by water vapor in the troposphere make VLBI observations at wavelengths significantly shorter than ~0.8 mm feasible from just a few extremely dry sites on Earth (such as the Chajnantor Plateau in Chile, Mauna Kea in Hawaii, and the South Pole), if the weather conditions are favorable.

Both of these limitations, the maximum baseline length and the limited availability of ground-based stations at high frequencies, can be overcome by placing all sub-millimeter VLBI telescopes in space. An additional advantage of such an approach is the possibility of obtaining better $(u, v)$-coverage when using an array of telescopes on freely flying satellites instead of an Earth-based array rotating as a solid body. Also, a dedicated fully space-based array will not be constrained by the need to apply for limited observing time at facilities that are in high demand for many other science cases. On the other hand, if the weather conditions are favorable during the permitted observation time, space-ground baselines could make use of the high sensitivity of ground stations like ALMA, and fill the $(u, v)$-plane on short (hourly) timescales. A space-space system observing together with ground stations when they are available could therefore be considered.

To date, three successful demonstrations of (partial) Space VLBI (SVLBI) systems have been achieved: TDRSS (1986–88), VSOP/HALCA (1997–2003) and RadioAstron (2011–2019) — see Ref. [18] and references therein. In all three cases, only one element of the SVLBI system operated in orbit. About a dozen other SVLBI studies considered various mission concepts, some of them with two or more space-borne elements, *e.g.*, iARISE[19] and the Chinese SVLBI initiative[20,21]. All three already implemented systems and multiple design studies of SVLBI concepts create a technological basis for prospective multi-element sub-mm SVLBI systems.

Another important technological contribution for prospective sub-mm SVLBI comes from developments of microwave systems for astrophysical studies such as in COBE[22], WMAP[23], Planck[24], and Herschel[25], and in Earth observation missions such as SMOS[26]. The latter is of special interest as an example of an aperture synthesis system in space.

A prospective sub-mm SVLBI system for studying SMBH shadows requires state-of-the-art space-qualified technologies and technology developments. These include precise baseline state vector determination, flight dynamics and altitude determination[27] (addresses all 3), high-rate data exchange between space-borne SVLBI elements as well as between them and an Earth-based data processing center[28,29], and ultra-fine synchronization of SVLBI elements. Some of the required technologies can rely on the existing developments in other applications, *e.g.* Ref. [28]. State-of-the-art technologies for the main reflector[24], and sub-mm receivers[25] are promising. The 4 m main reflector is bigger than in the ESA Planck. This is bigger than the 3.5 m in the ESA Herschel too. The single sideband (SSB) noise temperature aimed for in this paper is smaller than in the ESA Herschel HiFi instrument.

In this paper, we present the science goals and a concept of a multi-element sub-millimeter SVLBI system, provisionally called Event Horizon Imager (EHI), formed by satellites on near-circular polar medium Earth orbits[30]. We assess the best feasible performance of the key elements of the science instrument and supporting sub-systems in view of the latter's correspondence to the best characteristics achieved to date in other space-borne applications. We also simulate reconstructed images of the main science sources with the EHI.

## 2  Science Objectives

### 2.1  Imaging Black Holes Sgr A* and M87*

At sufficiently high frequencies ($\sim$ a few 100 GHz), the emission from the accretion flow is mostly optically thin and allows us to probe its geometry down to the smallest radii around the black hole. Theoretical simulations predict that this emission from the innermost accretion flow is bent around the black hole by its strong gravity, causing the appearance of a black hole "shadow"[1]. Due to the large angular sizes of Sgr A* and M87*, they are the most attractive targets for imaging a black hole shadow.

The Event Horizon Telescope (EHT) imaged the shadow of the black hole in the nucleus of the active radio galaxy Messier 87, M87*, as an asymmetric ring[8,9], and a model fitting led to a Schwarzschild shadow size of $39\pm4\,\mu$as[10,11]. The EHT aims to image also the SMBH shadow of the Sgr A*. This ground-based VLBI at the frequency of 230 GHz with an angular resolution of $\sim$25 $\mu$as to investigate directly the manifestation of general relativity effects in the immediate vicinity of a SMBH and to constrain models of the plasma flow around the black hole[2,31,32].

A next step in studying the SMBH shadow effect is distinguishing between geometries predicted by different theories of gravity. Based on images by Mizuno *et al.*[3] and Olivares *et al.*[33] comparing the size and shape of the black hole shadow, we estimate that an angular resolution better than $\sim$5 $\mu$as could potentially allow distinguishing between a Kerr black hole and a dilaton black hole, and between a black hole and a boson star. It should be noted, however, that this number was estimated based on two specific cases and may not hold under all circumstances. Another motivation for pushing towards a resolution of $\sim$5 $\mu$as is that parameters like the black hole spin may become measurable. The black hole spin not only affects the size and shape of the black hole shadow (at a level of $\lesssim$ 4%, see the previous section), but it also affects the orbits of the surrounding plasma, causing its appearance and dynamics to change. Van der Gucht[34] have assessed the possibility to recover black hole spin from theoretical source models using a machine learning approach. While the spin could not be recovered if the models were blurred to the EHT resolution ($\sim$25 $\mu$as), the correct spin was recovered from a range of five possible values in > 80% of the simulations of they were blurred to a resolution of 5 $\mu$as.

The EHI will test general relativity in a different regime than, *e.g.*, LIGO or VIRGO, which observe gravitational waves from merging black holes of tens of solar masses, and also from other objects like neu-tron stars. The EHI aims to observe supermassive black holes of millions to billions of solar masses.

Assuming a telescope with an Earth-size aperture (the limit for Earth-bound VLBI such as EHT), the angular resolution of $5\,\mu as$ requires an observing frequency above $1\,THz$. Terahertz-band interferometry on Earth-Earth baselines is challenging due to the strong atmospheric turbulence and absorption. Achieving the $5\,\mu as$ resolution from a low Earth orbit requires an observation frequency above $0.8\,THz$, which poses ambitious requirements to the interferometer coherence (Eq. (1) in Ref. [35]). At least three GNSS satellites should be visible from the EHI satellites to perform the accurate (carrier phase relative) navigation measurements. This limits the highest-possible altitude to $\sim 7500\,km$, at the initial GNSS visibility simulation[36]; the dedicated MEO navigation study[28] might refine this altitude and the number of GNSS transmitters in the common visibility of 2 navigation receivers by analysis of other navigation constellations (apart from GPS-only in Ref. [36]). Medium Earth orbits mitigate the observation frequency requirement to about $500\,GHz$ because of the baselines involved at altitude $\sim 7500\,km$. At $500\,GHz$, involving ground-based stations is challenging because of strong absorption, scintillation and scattering in the atmosphere. Space-space interferometry is not affected by these atmosphere-related issues, but it poses many engineering challenges.

An angular resolution of $\sim 5\,\mu as$ should be sufficient to image SMBH shadows in several other sources in addition to the prime candidates Sgr A* and M87*

(Table 1). As a secondary science case, the EHI will provide valuable information on the emission from the jets of other AGN (non-horizon) with unprecedented resolution in a hitherto unexplored frequency regime.

## 2.2 Simulated Sub-millimeter Emission from Sgr A* and M87*

In order to assess the required EHI system specification, we simulate its imaging performance in observations of the prime target sources. The Spectral Energy Distribution (SED) of emission from Sgr A* has a sub-mm peak[37], while SED of M87* emission peaks at $\sim 100\,GHz$ and slowly drops towards higher frequencies[38].

Out of the two primary targets, Sgr A* constitutes a more challenging case because of significantly more pronounced interstellar scattering effects as the line of sight lies in the Galactic plane. For the purpose of these simulations, we use as input a source brightness distribution consistent with a physical model of the direct SMBH environment[39,40].

As a benchmark observing frequency, we choose $690\,GHz$ following considerations on demonstrated capabilities of space hardware[24]. We use four General Relativistic Magnetohydrodynamics (GRMHD) model images of the accretion flow around Sgr A* as input for our simulated observations. These are the GRMHD models 16, 24, 31, and 39 from Ref. [39], ray-traced at $690\,GHz$. The models are considered to be plausible representations of the source structure of Sgr A* that fit the existing observational constraints. They represent physically different situations where the emission is dominated by an accretion disk (mo-

**Table 1** Horizon sizes and 230 GHz flux densities for several supermassive black holes which could be candidates for horizon-scale imaging with the EHI concept

| Source | Shadow size/$\mu as$ | 230 GHz flux density/Jy | R.A. to $\alpha/(°)$ | Dec. to $\alpha/(°)$ |
| --- | --- | --- | --- | --- |
| Sgr A* | 53 | 2.4 | 39.36 | 29 |
| M87* | 42±3 | 0.9 | 39.36 | 12.39 |
| IC 1459 | 9.2 | 0.26 | 27.23** | 36.46 |
| M84 | 9.1 | 0.13 | 40.8 | 12.89 |
| M104 | 5.7 | 0.25 | 37.06 | 11.62 |
| IC 4296 | 2.5 | 0.16 | 22.90 | 33.97 |

**Note** Values are from Ref. [7, 8]. The axis $\alpha$ is normal to the EHI plane (Figure 4, Subsection 3.2), and the angle with ** requires a 180° turn of the EHI dishes with respect to the EHI plane.

dels 16 and 31) or jet (models 24 and 39). They also consider different angles between the black hole spin axis and the line of sight (60° for models 16 and 24, and 30° for models 31 and 39). The models range from 1.1 to 4.1 Jy in total flux density. Collectively, these models sample the solution space within the boundaries set by our current knowledge of the source.

Interstellar scattering adds spurious refractive substructure to the observed image of Sgr A*[41−43]. This substructure is variable on timescales of about a day at 230 GHz and a few hours at 690 GHz. Over time, the average effect of scattering is a blurring of the source image by a scattering kernel that decreases in size with observing frequency. Simulated time-averaged images at 230 and 690 GHz, including the blurring effect from a Gaussian scattering kernel[44], are shown in Figure 1 for model 39. The extrapolated major axis of the scattering kernel (see $1.309 \pm 0.015\,\mathrm{mas\cdot cm^{-2}}$ in Ref. [45]) at 690 GHz is only $2\,\mu\mathrm{as}$[46], which is several times smaller than the maximum angular resolution that can be obtained with the investigated setup.

Additionally, we use the time-averaged M87* model from Ref. [40], which includes accelerated electrons to simulate emission from the relativistic jet. Electrons in the highly magnetized jet region are not expected to be in thermal equilibrium but are accelerated by, *e.g.*, magnetic reconnection events. The ray-traced model image is shown in Figure 2. The image was not scattered since interstellar scattering is considered to be negligible in the direction of M87*.

A frequency of 557 GHz was chosen here to align with the possible secondary science goal of imaging water in protoplanetary disks[47]. The EHI orbit radii are flexible and can be temporarily set to a shorter separation in order to image these disks on arcsecond scales. This secondary science goal is to be explored in more detail in future studies. Due to the flat continuum spectrum of M87*, the source morphology is expected to be similar at 557 and 690 GHz.

## 2.3   Variability of the Morphology of Sgr A* and M87* as a Function of Time

Sgr A* is variable on a timescale of minutes, which is much shorter than the time it will take to collect sufficient $(u, v)$-coverage for imaging. It has been shown[46,48] that making use of the linearity of the Fourier Transform, visibilities in the same $(u, v)$-cells obtained during different epochs with the EHT or EHI may be averaged to obtain an image of the average source structure, clearly showing the photon ring. The photon ring is a persistent feature that is unaffected by temporal variations in source morphology, instead it is completely determined by the lensing of light in the strong gravitational field close to the black hole.

EHI observations of a variable source have been simulated[46] using a running GRMHD movie[41]. This movie was looped every five hours and scattered with a moving scattering screen[42]. It was found that using complex visibilities the reconstructed images approached the average source structure, but the reconstructions deteriorate if one needs to rely on the bispectrum as the linearity does not hold for triple pro-

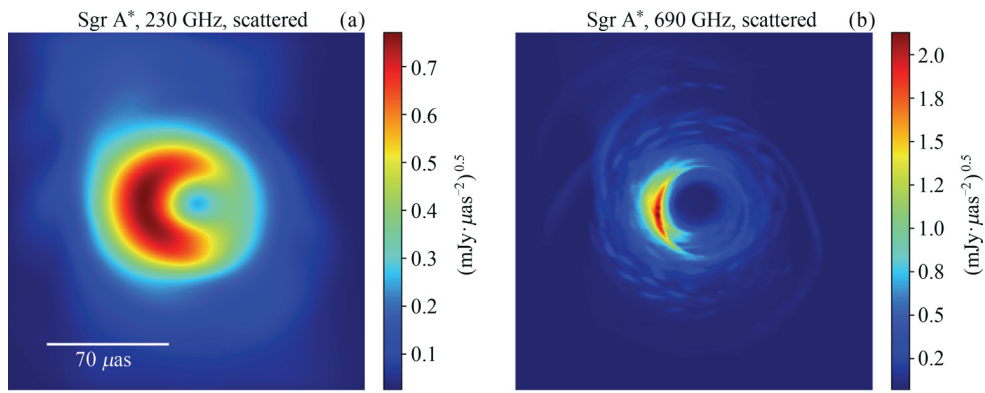

Fig. 1   Time-averaged simulated images of Sgr A* in the presence of interstellar scattering[39,45]. The 690 GHz image in the right panel offers the opportunity to characterize the shadow geometry down to an angular scale of ~5 μas, while the left image does not due to the lower instrumental resolution and the presence of interstellar scattering

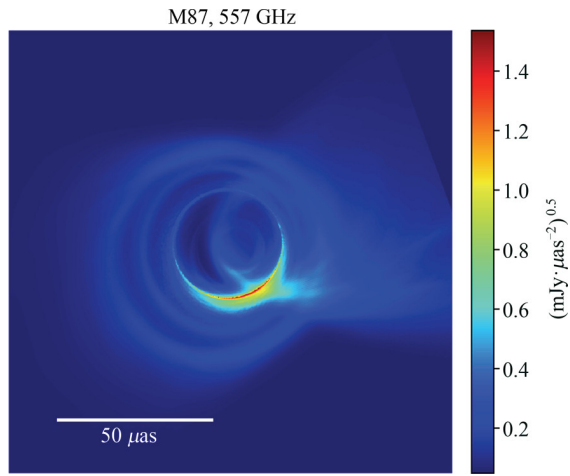

Fig. 2   Time-averaged M87* model[40] at 557 GHz. This
frequency was chosen to align with the frequency of a
water line detected in protoplanetary disks[47]. Imaging
water in protoplanetary disks might be considered as a
secondary science goal for the EHI. Due to the flat
continuum spectrum of M 87*, the source morphology is
expected to be similar over the range from 500 to
690 GHz, with its features dominated
by optically thin emission

ducts of visibilities. For M87*, the variability time-
scale is about 1500 times longer. Research is on-going
to see if the expected variability of M87* is slow eno-
ugh to allow a snapshot capability for EHI depend-
ing on the distance between the satellite orbits, which
gives a trade-off between the completion time of one
spiral iteration and the density of the $(u, v)$-coverage
(see Sections 3.2 and 4).

# 3   System Concept and Technical Requirements

## 3.1   Concept Overview

As discussed in Section 2, moving beyond the limits
on resolution imposed by Earth-bound VLBI requires
placing an interferometric instrument in orbit. The
aimed multi-frequency configuration of the EHI[49,50]
includes an observing band above 500 GHz as well
as two lower bands (which have been preliminari-
ly chosen to be at 230 GHz for comparison to EHT
and at 43 GHz to map jets, see Figure 6 in Ref. [51]).
The starting idea is that EHI satellites will initial-
ly function independently of the EHT telescopes, but
consideration is also being given to a hybrid system,
where the orbiting telescopes would be combined with

the ones on Earth, in particular at the lowest fre-
quency bands where atmospheric effects can still be
handled. The technical challenges to realize the hy-
brid system involve achieving coherent operation a-
mong all telescopes and data transfer to Earth, which
would lead to the use of optical links. Considering
radio regulations[52], a wide-enough frequency alloca-
tion band to downlink a radio-telescope's electrical
field data (IF), which is of the order of tens/ hun-
dreds Gbit·s$^{-1}$, does not exist, and hence an optical
link would be required. Using a hybrid system like
this could provide the possibility of creating moving
images of a black hole[53], and one might be able to
observe even more and also weaker sources.

The EHI concept is focused around multiple
satellites, ideally launched together in a single launch
vehicle, operating from circular Medium Earth Or-
bits (MEOs) at slightly different orbital radii. The
differential drift of the satellites with respect to each
other yields a wide range of sampled baseline lengths
and orientations, while the orbital plane can be cho-
sen such that the chosen sources can all be studied
with optimal $(u, v)$-coverage.

This section addresses the fundamental require-
ments pertaining to different aspects of EHI system
architecture.

## 3.2   Orbit Requirements

The angular resolution requirement for EHI ($\sim$5 $\mu$as)
can in principle be met by choosing an observing
frequency close to 500 GHz and an orbital radius of
$\sim$14 000 km. This choice of frequency, in combination
with this orbit size, provides the additional advantage
that the system is not limited in its angular resolu-
tion by interstellar scattering effects towards Sgr A*.
The specific choice of orbital radius is further con-
strained by the presence of the Van Allen radiation
belts, which limit the viable orbital radii to the re-
gion between $\sim$12 000 and $\sim$19 000 km, and the need
for GNSS positioning information. For successful cor-
relation, adequate knowledge of the baseline state vec-
tor and its time derivative will be important. The re-
quired availability of GNSS limits the maximum orbi-
tal radius for EHI satellites to be inside the outer Van
Allen belt (which spans orbital radii from 19 000 km
to 26 000 km).

Precise relative positioning in MEO using GNSS
is a subject that has been only modestly explored so far[54,55]. The expected real-time positioning accuracy for MEO is below 5 cm in position and 1 cm·s[-1] in velocity[56]. This is not yet at the required level for EHI, which needs a-posteriori (at image reconstruction *i.e.* the final) baseline vector knowledge to a fraction of the observing wavelength ($\leqslant 0.1$ mm for 500 GHz, see Ref. [36]). The sub-wavelength precision might be obtained as a bonus end-product of interferometric observations (from VLBI correlator) though the signal to noise ratio isn't enough for a detection. In order to achieve high-accuracy knowledge of both the baseline vector (the vector connecting the phase centers of the different telescopes in 3-D spatial coordinates) and its time derivatives, GNSS positioning measurements will need to be augmented with high precision optical metrology applied to EHI scales[57], or range-rate measurements over radio ISLs[36]. The ongoing investigation[36] into precise relative positioning in MEO to support science missions within NAV-ISP (the ESA Programme for Positioning Navigation and Timing), will allow us to refine our estimate for the accuracy of the achievable baseline state vector reconstruction.

We consider circular orbits to offer the most desirable geometry, as they will provide the most homogeneous $(u, v)$-coverage (see Figure 3). Furthermore, choosing an elliptical orbit[21] with an apogee closer to the GNSS shell will lower the quality of positioning information available for that segment of the orbit[56] or even eliminate GNSS coverage there entirely. The ongoing study of the precise relative positioning in MEO will valuate these reasons for the selection of orbit radii and orientation.

Regarding our choice for the orbital plane of the satellites, we aim for an orientation that offers good and persistent visibility of our two main science sources, Sgr A* and M87*. These sources are each located within $\sim30°$ from the celestial equator, making

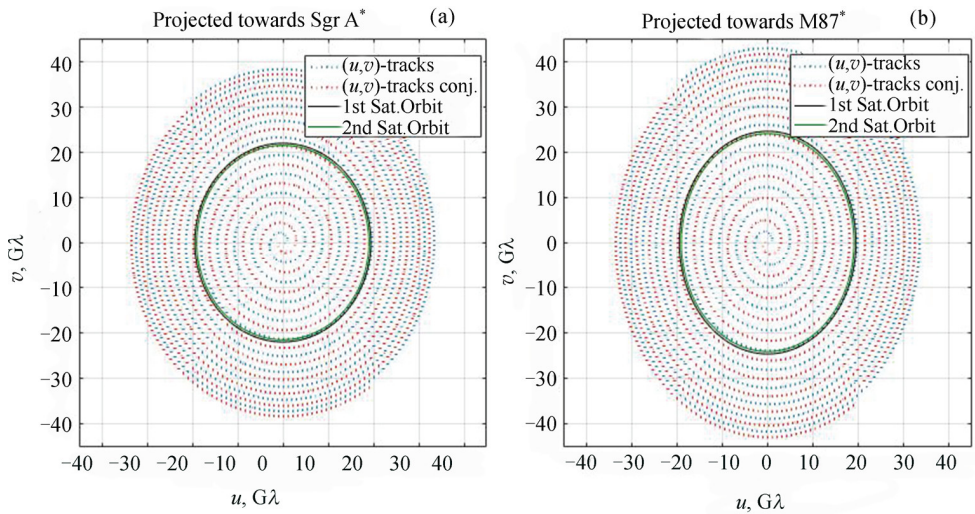

Fig. 3   Both the $(u, v)$-coverage and the orbits of two EHI satellites are projected towards Sgr A* and M87* in the left and right plots, respectively; the coverage with three satellites see Figure 1 in Ref. [46]. Because the satellites have slightly different orbital radii, their orbital periods are different, and their mutual separation (baseline) changes over time in both length and direction. This yields a spiral pattern in the $(u, v)$-plane. The maximum orbital altitude is $\sim7500$ km above the Earth's surface, which is the limit resulting from the initial GNSS visibility study. The difference in altitudes providing a spiral completion time of 1 month is 20 km, while in these figures the orbit altitude separation is 300 km to allow clearer visibility of the separate loops. The spiral opening starts when the inter-satellite distance is equal to their altitude difference. The spiral progresses further outwards until the inter-satellite link gets blocked by Earth's atmosphere, which here is considered to reach up to an altitude of about 100 km (the maximum central inter-satellite angle is $\sim\pi/\sqrt{2}$ rad.) Swapping the orbit altitudes allows changing from spiral opening to closing and vice versa. EHI calibration and commissioning benefit from swapping to the same altitudes, allowing for orbiting with a constant short baseline suitable for the highest correlated source flux density. These altitudes allow achieving the aimed angular resolution at 543 GHz, which is used in this figure. The consequences of adding a third satellite for purpose of forming closure phases are currently being investigated

high-inclination orbits the best choice for maximizing the extent of our $(u, v)$-coverage for either source. Simulations of ESA's Earth Observation missions suggested that the polar orbits can be known accurately[58,59]. The altitude increase (from LEO to MEO) mitigates the atmospheric drag for benefit of the relative navigation knowledge. The polar orbital plane maintains its orientation for the benefit of the main science goal (imaging major Sgr A*, M87* takes months) and a satellite on this inertial plane may maintain its orientation too. Furthermore, a polar circular MEO allows flexible orbit altitudes. When the inter-satellite link is blocked by the Earth (and the baseline is at its maximum), the orbits should be reversed to make the baseline shrink again. For an orbit altitudes separation of 3 km, such a maneuver can be done once every six months; such latency of 1 maneuver per 6 months is cheap and hence, preferred at a cost-effective mission.

For this reason, the polar orbital plane has been selected. The plane exhibits negligible precession and in this way, it is keeping its orientation, relative to the target sources, as stable as possible throughout the entire period of $(u, v)$-coverage accumulation, providing a stable evolution of $(u, v)$-coverage with time. Our choice of orbital plane is depicted in Figure 4. However, the ongoing research also considers the case of two orbits that are slightly slanted for the purpose of enhancing the relative navigation knowledge by using optical metrology in the out-of-plane direction.

### 3.3 Time of Baseline Presence in $(u, v)$-grid Cell

When we consider the way in which visibility measurements need to be gridded in the Fourier plane before their inverse transformation into the source image, we see that the choice of grid spacing is dictated by the overall angular size of the source on the sky – coarser gridding limits source image reconstruction to a smaller field of view. Connecting this to the way in which $(u, v)$-coverage is built up for EHI (see Figure 3), this means that limited time is spent in each $(u, v)$ grid cell per time the $(u, v)$-vector passes through that cell. This imposes an upper limit on integration time from the need to avoid combining visibilities from different $(u, v)$-cells into a single measurement, and it is an effect called $(u, v)$-smearing.

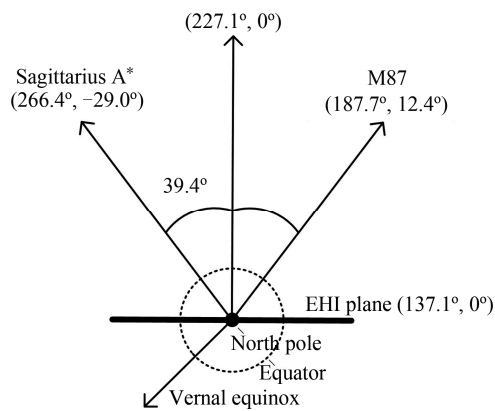

**Fig. 4** EHI orbital plane as seen from the north pole.

At this plane, the Right Ascension $(u, v)$-coverage projection narrowing (23%) is the same for both primary targets (Figure 3). The loss in projected baseline length in Declination is limited to 13% towards Sgr A*, and is negligible (2%) towards M87*. Baseline length losses towards other science sources (smaller than 10 μas in size in Table 1) occur because of this geometrical projection as well. A rotating this orbital plane ∼3.8° to (140.9°, 0°) enables to equalize the aperture projection towards both Sgr A* and M87 in future work

The dwell time per grid cell is shortest for the longest baselines. Following Eq. (2) of Ref. [46], the expression that gives the maximum integration time before $(u, v)$-smearing becomes an issue is given by

$$t_{\text{int}} < \frac{P}{2\pi b \theta_{\text{source}}}, \qquad (1)$$

where $t_{\text{int}}$ is the maximum integration time, $P$ is the orbital period of the satellites (and the cadence with which the $(u, v)$-vector circles the $(u, v)$-plane once), $b$ is the baseline length expressed in observing wavelengths (hence, dimensionless) and $\theta_{\text{source}}$ is the angular size of the source on the sky in radians. For a maximum baseline length of 50 Gλ, a source size of 150 μas and an orbital period of 4.5 hours we get a maximum integration time of 71 seconds if we wish to avoid $(u, v)$-smearing. This number scales inversely proportional to baseline length, allowing for longer time on shorter baselines. Namely, baseline lengths of an Earth diameter, 20 km, and 3 km allow a maximum integration time of 3 minutes, 1 day, and 1 week, respectively.

We assume that integration is constrained additionally ⩽ 7.5 min by $(u, v)$-arc length ⩽ 10°, at these orbit radii.

An orbit radii difference of 20 km is used in this paper (Figure 3). However, the EHI concept allows for manoeuvres between different orbital radii. A smaller difference could be useful for large-scale imaging of, *e.g.*, protoplanetary disks (Subsection 2.2). A larger difference (*e.g.* hundreds of kilometers) would decrease the time it takes to complete a $(u, v)$-spiral up to the longest available baselines, which allows for faster imaging with the trade-off of sparser $(u, v)$-coverage.

### 3.4  Imaging Approaches with Respect to Instrument Phase Stability Timescale

The type of imaging that is possible with the EHI depends on the system characteristics, most notably the coherence time scale.

A phase-stable interferometer (Case 1) can integrate during years (infinite instrument coherence timescale, the perfect knowledge of instrument positions). The correlator integration time, in this case, is limited to $(u, v)$-grid cell crossing time Eq. (1) but this provides no obstacle because images can be reconstructed by taking an inverse Fourier transform; there is no requirement to achieve any particular SNR within a $(u, v)$-grid cell crossing time (Subsecion 3.3). To provide the desired dynamic range, $(u, v)$-grid cells can be coherently averaged between numerous observation cycles (29 days each for an orbit separation of 21 km).

The next considered coherence timescale (Case 2) is one observation cycle. The instrumental phase, with contributions from the local oscillators, can be described in a delay and delay-rate (DnDR) model. The delay is the time difference of arrival of the science signal to the interferometer, and the delay-rate is its first time derivative. The instrumental phase can be measured using a calibration observation with the satellites orbiting on the same altitudes with a short baseline between them. The baseline will have a constant length and varying orientation, measuring the total correlated source flux density.

The derived DnDR can be fit for each observation cycle. As a result, $(u, v)$-grid cells can be coherently averaged between numerous calibrated observation cycles. Reversing the orbit altitudes allows closing the spiral back to the shortest baseline (Figure 3) and hence, the required phase stability time scale for such a mode is two spiral periods ($2 \times 29$ days). Another approach is to wait from the moment when the inter-satellite links get blocked by the Earth until the inner satellite catches up with the outer satellite and the inter-satellite link is not blocked anymore.

A coherence timescale (Case 3) shorter than the observation cycle (29 days) but still longer than the detection time (limited by the $(u, v)$-grid cell crossing time, see Subsection 3.3) would require having a significant SNR within a single $(u, v)$-grid cell crossing time, because this cell noise can not be averaged out anymore by multiple observation cycles. Also, there will be phase corruption between visibilities in different $(u, v)$-cells.

These phase offsets should either be calibrated (which is challenging due to the lack of unresolved sources that could be used for phase referencing at these high frequencies and long baselines), or a third satellite should be added to the system so that closure phases[60] can be used for imaging.

Closure phase is the phase of the triple product of visibilities on a triangle of baselines (bispectrum). Closure phases are immune to most station-based phase errors as long as the instrument coherence time allows the detection, or as long as the relative positioning information allows imaging without the detection stage. Hence, the closure phase information can be used for image reconstruction without imposing the technologically challenging requirement (Subsection 3.5) of a system maintaining phase coherence on longer timescales. Closure phases between different triangles are robust even if the individual visibility phases fluctuate. For these reasons, a three-satellite system is seriously considered for the EHI concept.

Finally, for a coherence time shorter than the $(u, v)$-grid cell crossing time (Case 4), the detection should be done within this coherence time. Phase corruption will appear between visibilities in different $(u, v)$-cells, similarly as in the former case above.

### 3.5  On Ultra-stable Oscillators

For sub-mm interferometry, frequency stability is of paramount importance to preserve phase coherence between the signals recorded at different stations right up to the stage of correlation. In VLBI, each station has its own (ultra-stable) oscillator, and the various Local Oscillators (LOs) are regularly disciplined using a lower-cadence shared signal. For instance, the EHT uses hydrogen masers at all stations, and through comparison with GPS over long timescales determines small fractional frequency offsets between the masers.

Frequency stability can be characterized by the Allan deviation, a figure of merit that characterizes the relative standard deviation in frequency of a clock referenced to a perfect reference clock over a range of different time scales. To maintain phase coherence, the Allan deviation for a given time interval multiplied by the observing frequency and the integration time should be limited to less than one radian (see Subsection 9.5.2 in Ref. [17])

$$2\pi f_c \tau \sigma_y(\tau) \leqslant 1, \qquad (2)$$

where $\sigma_y(\tau)$ is the Allan deviation, $f_c$ is the center frequency, and $\tau$ is the integration time in the interferometer, while a more exact expression for coherence time is in Ref. [35]. At $f_c = 500$ GHz, the required Eq. (2) Allan deviation of each ultra-stable oscillator is below $3.8 \times 10^{-15}$ (Hz/Hz), at integration time 60 s.

The heritage of ESA's ACES, existing and space-proven ultra-stable oscillators, includes the best-known space hydrogen maser[61,62]. This maser satisfies the requirement Eq. (2) on timescale up to 10 s, but it is 1.7 weaker than this requirement at 60 s.

The loss of interferometer coherence(Eq. (14), Eq. (18) in Ref. [35]) due to use of two such masers is within 16% on timescale $\leqslant 30$ s (35% at 60 s). The coherence loss on a timescale of 1–10 s is within 6%–8%. A better performing space-proven ultra-stable oscillator is not known.

The multi-pole linear ion trap frequency standard satisfies the requirement Eq. (2) on timescale up to 35 s, but it is 1.5 times weaker than this requirement on 60 s[63]. The loss of interferometer coherence(Eq. (18) in Ref. [35]) in case of use of two such frequency standards is within 16% on a timescale up to 35 s (29% at 60 s). The coherence loss on a timescale 1–10 s of is within 0.4%–4.2%.

The expected performance of the future Millimetron dedicated maser[64] (provisional launch in 2028) is 2 and 1.5 times better than of ACES maser on 10 s and 300 s, respectively. The smallest factor of 1.5 allows an EHI coherence time slightly above 60 s (Eq. (18) in Ref. [35]).

We assume that a future ultra-stable oscillator will allow a coherence time of 60 s minding that (i) the performance difference of RadioAstron and ACES masers is 1.5 times on timescale 1–1000 s, (ii) there is a difference between the goals in ACES, DSAC, and VLBI Millimetron/RadioAstron, and (iii) the EHI is a mission concept. This coherence time (60 s) is therefore used in our imaging and sensitivity calculations throughout this paper instead of the more conservative value of 30 s. This coherence time (60 s) is shorter than the $(u, v)$-grid cell crossing time (71 s–7.5 min, see Subsection 3.3), and hence the phase stability falls under case 4 in subsection 3.4. In order to upgrade from case 4 to case 3 or even 2, either a connected local oscillator architecture (see below) or a three-satellite system allowing for the measurement of closure phase is required.

The coherence time of 60 s constrains the allowed integration time (Subsection 1). However, it is longer than the atmospheric limit for ground-based observations, even as compared to an observation frequency band that is several times lower[64,65]. This is an asset of a space-space VLBI system at 500 GHz.

A connected-LO architecture may be required to improve the instrument coherence. One such scheme is as follows: master LO signals, generated on the different satellites, are inter-changed across each pair of satellites and locally mixed to generate a summed-frequency signal that is coherent across all of them. The inter-satellite paths used for the exchange of signals and LO components on each baseline are symmetrical and hence, each satellite has access to the same summed clock signal with the only difference arising from frequency variability on the signal propagation timescale between satellites (amounting to a maximum of ∼0.1 s). The achievable coherence time when using such a LO-sharing scheme is a subject of current research.

### 3.6 Telescope Requirements: Antenna Size, Bandwidth

As our aim is to keep the EHI cost-effective, we here consider that all satellites are launched together. The chosen observing frequency, of 500 GHz or higher, makes the option of a deployable antenna system challenging if practical, so we only consider the options for a static antenna here.

Such high-frequency band challenges the ultra-stable oscillators Eq. (2), constrains the integration time to 60 s, and motivates the use of the closure phase[60] to mitigate individual station-based phase errors.

The latter requires three science instruments embarked on separate satellites. The Ariane 6 fairing diameter allows up to 4.0 m main reflectors, using a Cassegrain configuration (Figure 5).

The system sensitivity will be accordingly limited: using an aperture diameter of 4.0 m, an estimated system temperature of 150 K, and a total system efficiency as described in Ref. [46] we get a System Equivalent Flux Density (SEFD) of $6.8 \times 10^4$ Jy for each of the antenna systems. For comparison, the median SEFD values for the EHT stations during the 2017 observations range from 74 for Atacama Large Millimeter Array to $1.9 \times 10^4$ Jy for the South Pole Telescope[16].

The low correlated flux density expected from our main sources on long baselines poses a challenging sensitivity requirement and hence, the widest possible observing bandwidth is desired. Existing Earth-bound telescopes operating at mm bands process intermediate frequency bandwidths of 4–8 GHz[66–68]. The latest developments in mm and sub-mm receiver technology allow an intermediate frequency bandwidth up to 22 GHz[69,70]. The widest-possible bandwidth is desired to recover the sensitivity. However, the bandwidth poses requirements on the real-time knowledge of the DnDR model[71] and the achievable values are expected from Ref. [36].

For the EHI, we take a processed bandwidth of 5 GHz. Assuming that the system observes two orthogonal polarisations at 2-bit Nyquist sampling, we get a corresponding aggregate raw data rate of $5 \times 10^9 \times 2 \times 2 \times 2 = 40$ Gbit·s$^{-1}$ to be transmitted between satellites.

The highest data rates for any space communication link that have so far been demonstrated are 5.6 Gbit·s$^{-1}$ and 1.8 Gbit·s$^{-1}$ over distances of 5100 km (LEO–LEO) and 45 000 km (LEO–GEO), respectively[72,73]. The next generation of the optical space communication terminals aim to reach a data rate of 1.8 Gbit·s$^{-1}$ over distances up to $\sim$80 000 km[74].

Ongoing developments will push this figure to higher values. ESA's program Advanced Research in Telecommunications Systems (ARTES) includes a number of activities (with Synopta-CH, AirbusDS-Fr, TNO-NL) to develop terabit links for future communication satellites using Wavelength Division Multiplexing (WDM) technology with the aimed capacity of 10 Gbit·s$^{-1}$ per carrier wavelength[28]. A new project High Throughput Optical Network (Hy-dRON), also under preparation by ESA[28], aims to demonstrate 0.1–1 Tbit·s$^{-1}$ bidirectional space optical link capability, among other goals, and the planned demo-phase is in 2023–2025. The above mentioned next generation of optical space communication terminals (under development in TESAT) is one of the major developments[74]. Ranging can be implemented in optical communication terminals addressed in Ref. [74].

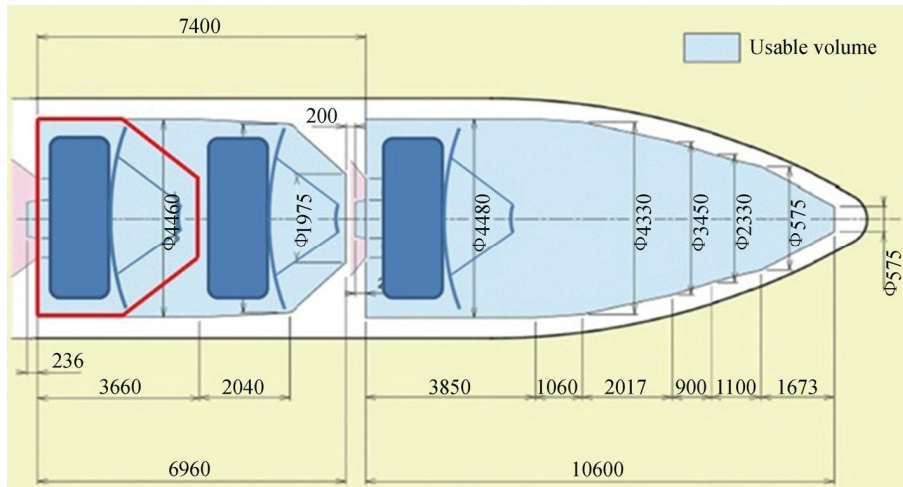

Fig. 5   Accommodation of EHI satellites in Ariane 6

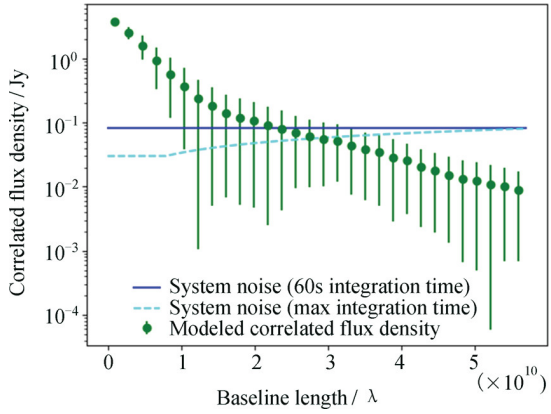

Fig. 6   Plotted in green: expected correlated flux density as a function of baseline length for one of the models for Sgr A* (model 24), using radial bins in the $(u, v)$-plane. The continuous blue line shows the system noise floor for an integration time of 60 s. In cyan, the sensitivity following from the maximum integration time allowed by $(u, v)$-smearing is shown with a maximum allowed integration time of 7.5 min

## 3.7   Sensitivity

To estimate the visibility noise for a given observing bandwidth and integration time, we use the standard[17] expression:

$$\sigma = \frac{1}{0.88} \sqrt{\frac{S_1 S_2}{2B\tau}}, \qquad (3)$$

where $\sigma$ indicates the system noise level in Jy, $1/0.88$ is the standard Van Vleck 2-bit quantisation correction factor, terms $S_1$ and $S_2$ indicate the system equivalent flux densities (in Jy) of the two stations, $B$ is the observing bandwidth in Hz, and $\tau$ is the integration time in seconds.

Given the choice of orbital plane and orbital radii, we will get maximum projected baseline lengths towards our prime sources in the range from 47 Gλ to 53 Gλ, at 690 GHz. The expected correlated flux density for Sgr A* on these long baselines, obtained from our source models[46], is lower than 30 mJy (Figure 6). The integration time required to achieve a 7-sigma detection on a baseline of 50 Gλ is 6 h, while EHI baseline vector crosses the corresponding $(u, v)$-grid cell on 71 s (see Subsection 3.3) and hence, obtaining a detection at the longest baselines is not feasible.

Baseline shortening allows longer integration time Eq. (1), but limits the angular resolution and hence, prohibits the principal science goal. In detail, the achievable baseline length allows 19 μas instead of 5 μas towards Sgr A*, at 71 s (see 60 s in Subsection 3.5).   More challenging performance is required for the target resolution (5 μas) at this 7-sigma detection threshold. For example, a receiver noise temperature 50K and observation bandwidth 10 GHz would still require a 20 m main reflector which isn't feasible at surface accuracy below 10 μm and at a cost-efficient mission.  Narrowing the DnDR search window sizes is promising to decrease the detection threshold and other source models, see models 16, 31, 39 and Figure 3 in Ref. [46], feature several times lower flux densities and hence, indicate additional challenge for this development avenue.

The relative positioning is little explored on Medium Earth Orbits, while the perfect knowledge of DnDR model (instrument baseline state vector, translation between phase centers of antennas, difference in local oscillators) allows image reconstruction without any detection required beforehand, as demonstrated in the ESA SMOS[26,75]. In case the coherence timescale is longer than the observation cycle, the uncorrelated noise can get averaged out by a number of observation sessions as mentioned in Subsection 2.3 in Ref. [46].

The EHI concept[50,56] aims to relax the detection requirements by the involvement of relative navigation and connection of ultra-stable oscillators. As above mentioned, a detection might not even be needed at all, if the orbit determination is accurate enough (e.g. with a 3D baseline state vector accuracy of a fraction of the observing wavelength, see Subsection 3.2). Refined knowledge of the baseline state vector, available a posteriori (as reconstructed from an advanced relative orbital propagation model), potentially allows us to reduce the DnDR search window for fringe fitting, correspondingly improving the detection threshold. Once the detection is successful, further improvement in visibility SNR may be achieved by coherent averaging of visibilities within each $(u, v)$ grid cell provided that the coherence time is not shorter than grid cell crossing time. Alternatively, bispectra[60] may be formed from the visibilities and averaged over multiple epochs.

Assuming that the navigation study disables blind selection of the data bin, detections for Sgr A* are limited to the shorter baselines, this does not hold for other sources (see e.g. Table 1). All of the AGN sources besides Sgr A* and M87* are expected to be more compact in terms of their size on the sky, and several of them are actually brighter in total flux density (*e.g.*, 3C273 and 3C279). This means that we can expect higher correlated flux density levels on long baselines for several of these sources, making detections possible further out in the $(u, v)$-plane and offering a higher angular resolution for those sources.

### 3.8 Correlation Scheme

A downlink of electric field signals would allow versatile re-processing and the required downlink data rate is $40 \, \text{Gbit·s}^{-1}$ (see Subsection 3.6). Such data stream can not rely on radio frequency allocation bands for the downlink of space research data from MEO (those feature width up to 1.5 GHz, see Ref. [76]) in S-, X-, and Ka-bands, and on-ground stations supporting those. The W-band allocation (within 74–84 GHz) is not capable enough too and moreover, the modulator technology for such a wide-band is challenging[29]. For these reasons, the electric field data must be exchanged between satellites and correlated on one of them.

In order to cross-correlate the sampled signals from the different satellites with the right delay model, accurate knowledge of the 3D baseline vector and its time derivative(s) is required. Perfect knowledge of this baseline geometry would allow us to blindly pick the correct correlation peak from the correlated products at all times. In reality, this will not be the case: residual errors in satellite positions and velocities mean that we will need to use a DnDR search window.

We adopt a usual VLBI processing architecture wherein the first stage computes the cross-correlation across a range of values in DnDR over relatively short accumulation periods using an a-priori delay model, and the second stage solves for the residual DnDR (fringe finding), thus allowing integrating over longer time scales, see Subsection 9.2 in Ref. [17]. In this proposed scheme the first stage of processing is to be done on-board the satellite stations using a delay model based on real-time orbital information. The data product of this stage is then transmitted to the ground station, where the second stage proceeds using a more accurate delay model based on orbital reconstruction which allows coherent addition over the much longer time scales required to yield sufficient SNR.

The on-board (first stage) accumulation period is limited by the accuracy of the real-time delay model, see Eq. (3) in Ref. [71]. We assume in Subsection 3.9 that real-time positioning accuracies (1-sigma) of delay-rate and acceleration in 3D relative coordinate system are smaller than $\sim 60 \, \text{ps·s}^{-1}$ ($1.7 \, \text{m·s}^{-1}$ baseline velocity) and $\sim 80 \, \text{ps·s}^{-2}$ ($2.5 \, \text{cm·s}^{-2}$ baseline acceleration), respectively. This allows an accumulation period of 0.3 s, at a bandwidth of 5 GHz (0.1 ns Nyquist sample rate).

At an observed frequency of 500 GHz, an on-board integration time of 0.3 s dictates that the delay-rate resolution be $6.67 \, \text{ps·s}^{-1}$ and a delay-rate window of 34 samples ($\pm 3.4 \, \text{cm·s}^{-1}$) is required (113 samples at a higher delay-rate resolution $2 \, \text{ps·s}^{-1}$). Assuming centimeter-level accuracy in the projected baseline length, a delay window of the order of a few samples is sufficient for the observation signal coherence length down to 6 cm, at bandwidth 5 GHz. The length of the delay window should allow also for the spectral resolution required to calibrate for the instrumental bandpass shape, and for digital sideband separation, see *e.g.* Ref. [77]. A coarse spectral resolution allows narrowing the search window to reduce the downlink data rate. A delay window of 40 samples enables calibrating for 500 MHz-scale (10% bandwidth) spectral features in a 5 GHz band width. This delay window width is narrowed using Hermitian symmetry at the output of the Fourier transform, and keeping both unique real-valued numbers on 0 Hz and the Nyquist frequency. The search window size number is $21 \times 34 = 714$ ($21 \times 113 = 2373$ at $2 \, \text{ps·s}^{-1}$). The output data rate to be transmitted to the ground in this scenario is of $3 \times 2 \times 4 \times 21 \times 34 \times 32 = 550 \, \text{kbit·s}^{-1}$ per 5 GHz band width, accounting for baselines between 3 satellites (or 3 observation frequencies), 4 polarization, 2 quadrature products, and 32-bit data ($1.8 \, \text{Mbit·s}^{-1}$ at $2 \, \text{ps·s}^{-1}$).

This data rate is $\sim 438\,000$ times smaller than it would be in the case of electric field signals ($132\,000$ times at $2 \, \text{ps·s}^{-1}$). This data rate fits any of the aforementioned S-, X-, Ka-band allocations and allows EHI to benefit from ground stations supporting those. This is an asset of being able to perform the first stage of processing on the fly.

In case a wider downlink is needed in future, the Shannon–Hartley theorem sets the upper bound for the (Additive White Gaussian point-to-point) communication link capacity equal to ~5.2 Gbit·s$^{-1}$, at a link budget of (SNR) 10 dB and within this allocated frequency band width of 1.5 GHz (the margin's upper bound is 3 orders of magnitude between 5.2 Gbit·s$^{-1}$ and 550 kbit·s$^{-1}$).

The second stage of this processing scheme is dedicated to adjusting and averaging-up (within the coherence time) the downlinked DnDR search windows according to the positioning knowledge available after precise relative orbit reconstruction on the ground. The latter adjustment includes a correction of the DnDR model applied on-fly using its refined values available on-ground.

The alternative case, that the final orbit reconstruction does not allow this selection, is not addressed in detail because the navigation study result might allow a sufficiently accurate knowledge of the DnDR model. Namely, it may be assumed that in this alternative case of detection (a) a false alarm rate of 1 in 100 is acceptable and (b) the correlator output values are normally distributed and hence, a number above a threshold of 2.58-sigma might be a signal. This threshold level constraints the longest baseline (Figure 6) and hence, precludes the aimed angular resolution. The on-going navigation study[36] may result in an orbit reconstruction accuracy that allows for image reconstruction with a lower SNR per $(u, v)$-point.

For orbits that meet the requirements described above, the maximum digital delay required to align the voltage data stream digitized on one satellite with the data stream received over the ISL from another satellite is of the order of 100 ms, or a few Gbits per 5 GHz band width. The maximum rate of change in the delay is of the order of 10 μs·s$^{-1}$, or equivalently for integer sample delay, an increment of one per hundred thousand. Limiting the loss due to sub-sample delay errors[17] to below 1% can be achieved by applying a spectral-domain correction in an FX-architecture, which would require updating once every ~10 FFT windows. In order to limit fringe rotation loss[17] to an acceptable level requires updating the fringe rotation phase at a rate of up to a few ev-ery ~100 samples, and partial fringe rotation would likely need to be done in the analog domain.

## 3.9 Relative Positioning

The feasibility of the correlation scheme described above is dependent on having knowledge (not control) of the inter-satellite baseline vectors with centimeter-level accuracy in orbit, and sub-mm level accuracy on ground. A candidate set of Altitude and Orbit Determination Systems (AODS) is in Subsection 2.3 of Ref. [36], and the ongoing industrial study is dedicated to refine it; the translation from the measurements with a refined set of AODS, GNSS receivers, and inter-satellite metrology to the satellites' centers of masses (and the need or not for it) and, finally, to the phase center of the interferometer is the the subject of the latter study, as well as the time lag between science and navigation measurements. It is unlikely that this sub-mm level of accuracy can be reached in real-time in 3D spatial coordinates, which means that at the time of on-board correlation the relative positioning accuracy just needs to be small enough (see both cm-level and downlink *margin* in Subsection 3.8) to provide DnDR search windows of a reasonable and limited width. With a-posteriori relative orbit determination accuracy, the correct DnDR bins may then be selected on the ground. Note that, if the quality of a-posteriori relative positioning (including phase center tracking) is sufficiently high, the correct bin in the DnDR search window may be picked "blindly" without needing to obtain a detection first. This is a very important point, as it potentially allows us to use longer effective integration times for (complex visibilities or) bispectra by coherently adding them together over multiple epochs.

At this point, we can consider specific numbers for the quality of relative positioning needed for EHI. As derived in the preceding section, to successfully perform cross-correlation with an integration period of 0.3 s in space the real-time delay-rate and acceleration need to be known to a precision of 1.7 cm·s$^{-1}$ and 2.5 cm·s$^{-2}$, respectively. For the second stage of fringe finding, done on Earth after the correlation products have been downloaded, a different (better) a-posteriori baseline knowledge is needed. In order to have the capability to blindly pick DnDR solutions using a-posteriori orbit reconstruction information, the baseline vector knowledge needs to be known to within a fraction of the central observing wavelength. Specifically, EHI beam directivity diminishes by 0.5 (3 dB) at a baseline vector knowledge accuracy (1-sigma in 3D relative coordinates) of $\sqrt{2}\lambda/6 = 0.14$ mm, at an observing frequency of 500 GHz, see Figure 1 in Ref. [36]. This assumes that other sources of errors in DnDR model are absent and as a result, delay, delay-rate and acceleration accuracies below about $60\,\mu$m, $80\,\mu$m·s$^{-1}$ and $120\,\mu$m·s$^{-2}$ are necessary for this stage. We will now consider how close the current capabilities for relative positioning between satellites come to these numbers.

### 3.9.1 Demonstrated Positioning performance

In LEO, relative positioning has been demonstrated with accuracies down to millimeter-scale in the Gravity Recovery and Climate Experiment (GRACE), and the TerraSAR-X add-on for the Digital Elevation Measurement (TanDEM-X) mission[78–81].

Relative positioning accuracy down to mm/sub-mm scales has been achieved in GRACE by means of ranging instruments, providing satellite-to-satellite distance tracking with an accuracy at the $\mu$m-level in K/Ka-band[78]. Optical inter-satellite metrology will be applied in EHI[56], which is expected to yield even higher ranging accuracy.

Although absolute satellite positioning is not immediately relevant for EHI as we are primarily interested in relative satellite positioning for the baseline geometry, the performance shown so far for absolute satellite positioning in various missions provides a reference level of accuracy upon which we can base our expectations for EHI.

The absolute positioning accuracy achieved in the ESA Copernicus satellites is < 2 cm with 0.5 cm residuals[58,59,82]. For Sentinel-3B, absolute positioning accuracies are about 1.25 cm, 0.7 cm, and 0.5 cm with latencies of 30 min, 1.5 days, and 25 days, respectively[58].

For the GRACE, TerraSAR-X/TanDEM-X, and Swarm missions, the differential orbit determination accuracy expressed in Earth-Centered Earth-Fixed coordinates is below 1.4 cm (1-sigma in 3D coordinates)[78]. Although the relative positioning knowledge in Prototype Research Instruments and Space Mission technology Advancement (PRISMA) mission falls far short of the precision required for EHI (inter-satellite metrology uses decimeter waves versus micrometer waves in EHI), the ratio between absolute and relative positioning accuracies for that mission is about 38[83]. Relative positioning accuracies of Proba-3 (Project for On-Board Autonomy-3) haven't been demonstrated yet (to be launched in 2020).

### 3.9.2 Availability of GPS Transmitters

Unlike for satellites in LEO, precise satellite positioning in MEO has not been studied in detail to date. This somewhat limits useful extrapolation of the relative positioning performance from LEO to MEO, which is why a dedicated study on precise navigation in MEO is currently being undertaken[36]. We do however have certain expectations on what precision should be attainable, which we will discuss here. A number of important conditions are different for MEO versus LEO. The following factors simplify the case for MEO orbit propagation: compared to LEO, satellites in MEO are practically free from atmospheric drag, their orbital dynamics feature a smaller influence from higher-order gravity modes, and ionospheric effects on signal propagation are also avoided. The attainable performance depends also on the number of GNSS transmitters in view of the EHI navigation receivers (which is smaller for MEO than for LEO due to longer baselines and closer proximity to the GNSS shell), and on inter-satellite metrology performance.

This difference in GNSS availability has been quantified in a previous study[36]. The simulated number of GPS transmitters in the common field of view of EHI navigation receivers is up to 10 at the shortest baseline lengths and decays as the baseline increases. The mean duration of GPS satellites in common visibility ranges from 7–22 minutes, and the maximum time in common visibility ranges from 21–154 minutes. Precise navigation during time intervals outside the availability of GPS satellites could be supported by data from inertial navigation sensors and inter-satellite metrology[50,56]. Taking into consideration all of the 6 available GNSS constellations instead of the only GPS is expected to increase the visibility of GNSS satellites by more than 4 times.

### 3.9.3 Demonstrated Performance of Inter-satellite Metrology

EHI maintains the permanent connection between optical terminals and hence, microwave inter-satellite metrology[36] becomes impractical. The optical inter-satellite link concept offers EHI a data rate of $40\,\mathrm{Gbit\cdot s^{-1}}$ and simultaneously a ranging accuracy within $0.1$–$4\,\mathrm{mm}$, depending on the bit interpolation applied, following Ref. [74].

Laser ranging accuracy $5\,\mathrm{cm}$ (1-sigma) has been required on distances up to $42\,000\,\mathrm{km}$ and demonstrated on the ground[84,85]. Further industrial work on that is supported by the ESA TRP program and SkyLight element of the ESA ARTES program[28]. Link budget optimization is a promising way to enhance metrology performance on the longest baselines[86,87].

EHI benefits from development of technologies for optical terminals and establishing the links in these ESA programs.

A version of inter-satellite metrology (the science payload) at gravimetry missions could refine the DnDR model at EHI. Both inter-satellite range and range-rate may be measured on EHI baselines, and blended together with AODS and carrier-phase relative GNSS data, in parallel to science observations. The exact timescale it takes to minimise the error of this real-time navigation knowledge to a cm-level is under investigation by the industrial study. The aimed inter-satellite distance knowledge accuracy is down to $\sim 0.1\,\mu\mathrm{m}$ in gravimetry missions (CHAMP, GRACE, GRACE follow-on (GRACE-FO), GOCE)[88,89]. The above-mentioned microwave instrument on-board GRACE features $\mu\mathrm{m}$-level accuracy[78]. The first-ever laser ranging interferometer demonstrator (on-board GRACE-FO mission, launched in 2018) showed in space the biased range measurements (similar to the primary ranging instrument based on microwaves) with noise at a level of $1\,\mathrm{nm\cdot Hz^{-0.5}}$ at Fourier frequencies above $100\,\mathrm{mHz}$[90,91]; the LRI is LISA demonstrator. Due to $10\,000$ times shorter wavelengths between K/Ka-band and optical instruments, the aimed measurement accuracy in GRACE-FO is $80\,\mathrm{nm/\cdot Hz^{-0.5}}$ in the measurement band $2$–$100\,\mathrm{mHz}$[89]. The NGGM laser metrology accuracy goal is further smaller. The latter band is similar in all gravimetry missions, because it is driven by gravimetry science needs and orbital scenario on one hand ($100\,\mathrm{mHz}$), and by intersection between measurement noises of ranging instrument and atmospheric drag interferometer ($2\,\mathrm{mHz}$) – on another hand. In EHI, such band is of $6\,\mathrm{Hz}$–$7.5\,\mathrm{kHz}$ given by the inter-satellite round-trip. Much higher frequency offsets are promising to operate in lower frequency noise.

The EHI longest baseline is 2 orders of magnitude greater than the such in these gravimetry missions but both the lower frequency noise and completed studies for laser ranging over $42\,000\,\mathrm{km}$ distances are promising. The aforementioned studies could be merged to develop for EHI the optical communication terminal that features the required range and range-rate capabilities too. The feasibility of this should be covered in the dedicated study. The employment of lower-frequency observation band data (the candidate $43\,\mathrm{GHz}$ band for AGN jet imaging) to support relative positioning is currently under early investigation.

The spatial offset between instrument antenna, navigation antenna and inter-satellite link antenna may be designed to remain stable after launch. The calibration orbiting phase could provide the bias which remains between them after launch.

### 3.9.4 Expected Relative Positioning Accuracy in EHI

Extrapolating LEO performance to MEO, we do not expect relative positioning to be worse than absolute positioning. With respect to the above-mentioned (see Subsection 3.9 in Ref. [50]) and the GPS-visibility simulation[36], the expected real-time relative positioning accuracy for EHI is below $2\,\mathrm{cm}$ ($2\,\mathrm{cm\cdot s^{-1}}$, and $3\,\mathrm{cm\cdot s^{-2}}$), and a-posteriori this expectation improves to a value in order of $50$–$100\,\mu\mathrm{m}$ ($50$–$100\,\mu\mathrm{m\cdot s^{-1}}$, and $70$–$140\,\mu\mathrm{m\cdot s^{-2}}$). Currently, a precise relative positioning study[36] is underway by the ESA to provide a solid expectation for accuracy in MEO.

An imperfect a-posteriori knowledge of the DnDR reduces the coherence and prohibits an operation in the phase-stable interferometer mode (Case 1, Subsection 3.4). The expected knowledge of the DnDR model ($(50+100)/2 = 75\,\mu\mathrm{m}$, $(50+100)/2 = 75\,\mu\mathrm{m\cdot s^{-1}}$, and $(70+140)/2 = 95\,\mu\mathrm{m\cdot s^{-2}}$) starts to get into case 2 (Subsection 3.4). However, these accuracies still correspond to a phase error std of $47°$ at the observing frequency ($500\,\mathrm{GHz}$), which would allow coherent averaging of data to build up signal to noise but would introduce a phase error limiting the image the dynamic range by a (loss) factor of 2.4, see 3.8 dB at $\lambda/5.4$ in Figure 1 of Ref. [36]. A mitigation of this loss by factor of 2 (from efficiency $1/2.4 \simeq 0.4$ to $\sim 0.8$) would require an a-posteri position accuracy knowledge improvement by a factor of 2.3, which is very challenging. A true phase stable interferometer (Case 1, Subsection 3.4) would allow for a lossless maximum image dynamic range limited by the thermal noise only. However, that would likely require a-posteri position accuracies to improve by a factor of 4.5, which is challenging.

### 3.10    Alternative SVLBI Concepts

The concept described in the previous sections is challenging because of the high frequencies and long baselines involved. One could also consider alternative concepts. For instance, Palumbo *et al.* [53] consider launching satellites into Low Earth Orbits (LEOs) to observe in conjunction with the EHT at 230 and 345 GHz. This setup allows for a fast $(u, v)$-plane filling rate so that the dynamics of Sgr A* may be captured, allowing for the reconstruction of movies of the surrounding plasma. While we consider this science case worthwhile, it will not give the $\sim 5\,\mu$as angular resolution aimed for in this project (Section 2). Also, the observations will be affected by the atmosphere and interstellar scattering, the $(u, v)$-coverage would not be as dense and uniform as for the EHI, and one would need to apply for time at Earth-based telescopes.

Another alternative is described by Fish *et al.* [92], who consider launching satellites into Geosynchronous Orbits (GEOs) observing in conjunction with the EHT and satellites in LEOs and MEOs at 230 GHz. This concept would allow for an angular resolution that is comparable to that of the EHI, but the $(u, v)$-coverage would be significantly sparser and one would again need to rely on Earth-based telescopes and deal with the atmosphere. About a half dozen satellites would be needed to produce high-fidelity images with this concept.

Finally, studies have been carried out simulating the effect of participation of the Millimetron satellite in EHT observations [64]. Although the imaging resolution will be improved and dynamical imaging reconstructions may be possible, the $(u, v)$-coverage will be relatively sparse and the angular resolution will not meet our science objectives.

## 4    Imaging Simulations

In order to estimate the image quality that can be obtained, we present imaging simulations of general relativistic magnetohydrodynamics (GRMHD) models of the accretion flow in Sgr A* and M87* using the space-space interferometer concept described in this paper.

Following Ref. [46], model visibilities were calculated and the noise was added using the system parameters assumed in Section 2.2. The eht-imaging library [93,94] was used to calculate the model visibilities. For the $(u, v)$-coverage, we assumed an integration time per measurement equal to the maximum integration time from Eq. (1), with the additional constraint that the $(u, v)$-arcs swept out per integration should not exceed 10 degrees. The satellites were placed in orbits with radii of 13 892 and 13 913 km. For the simulations with closure phase, a third satellite was placed at a radius of 13 899 km.

Complex visibilities can be used for image reconstruction if the interferometer is phase-stable during at least one iteration of the spiral (29 days), which is technically challenging (see 60 s in Subsection 3.5). We present images for two different cases taking into account ongoing work on LO-sharing scheme, namely the phase-stable interferometer and the three-satellite system measuring closure phase.

Simulations we are presenting here assume a perpendicular orientation of the orbital plane with respect to the line of sight, and a good image can still be reconstructed if the baseline lengths divided by $\sim 2$ in one direction, see Figure 11 in Ref. [46].

### 4.1    Simulated Images of Sgr A* for Both Cases of the Phase-stable Interferometer, and Closure Phase

The input model for Sgr A* is model 39 (scattered) from Ref. [39] at 690 GHz (Figure 1, Section 2.2).

The top row of Figure 7 contains the model images reconstructed under the assumption that the two-satellite configuration of the EHI is limited by its expected sensitivity only, *i.e.* all other system parameters (including coherence and positioning) are excellent. The complex visibilities were gridded with a grid cell size set by the model image field of view of 210 $\mu$as and averaged over multiple iterations (months) of the spiral. The images were reconstructed by perfor-ming an inverse Fourier Transform on the gridded visibilities directly. This simple imaging procedure is possible here due to the dense and isotropic $(u,v)$-coverage. The reconstructed image shows significant improvement as the total integration time increases from 1 to 24 months and the cell-averaged visibility SNR accumulates. Figure 8 shows the corresponding SNR increase of the gridded visibilities.

The reconstructions in the bottom row if Figure 7 were made with the eht-imaging library[93,94] using bispectrum information only. The use of the bi-spectrum offers a way to coherently average measurements taken over long timescales - even if the visibility measurements themselves show significant phase wander. The integration time per bispectrum measurement was set to the $(u,v)$-smearing limit Eq. (1). For the multi-epoch reconstructions, the complex bispectrum was averaged rather than the complex visibilities. The dish diameter was set to 4.0 m rather than 4.4 m so that three satellites would fit in an Ariane 6 fairing, and l ess information (bispectrum instead of complex visibilities) was used to reconstruct

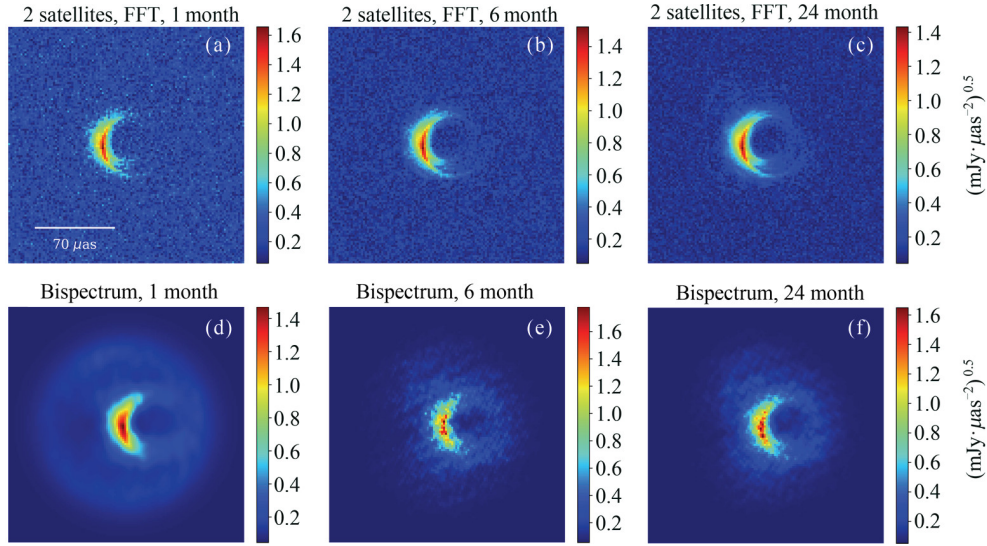

Fig. 7   Image reconstructions of the time-averaged Sgr A* GRMHD model 39 at 690 GHz from Moscibrodzka *et al.*[39] using two satellites with $(u,v)$-gridding assuming a system behaving like a phase-stable connected interferometer (top row), and three satellites using closure phases and visibility amplitudes (bottom row), with total integration times of 1, 6, and 24 months (left to right). Note that the system noise for the images in the bottom row is ∼17% higher than that for the top row (see text). Simulations are from Ref. [46]. Credit: F. Roelofs, A&A, 625, A124, 2019, reproduced with permission © ESO

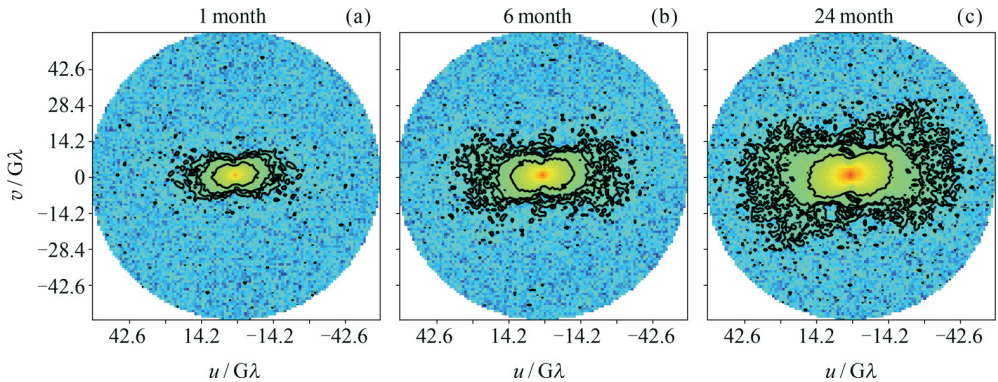

Fig. 8   SNR map of the gridded visibilities of model 39 (scattered) at 690 GHz after integrating for 1, 6, and 24 months (left to right), with a reflector diameter of 4.4 meters. Contours indicate the points with an SNR of 3, 5, 7, and 20. Figure is from Ref. [46]. Credit: F. Roelofs, A&A, 625, A124, 2019, reproduced with permission © ESO

the images. These factors together likely cause the image quality to be worse than for the gridded complex visibilities (top row).

For comparison, Figure 9 shows an image reconstruction made from a simulated ground-based observation with the EHT stations that participated in the 2017 EHT observing campaign, plus the NOEMA interferometer on Plateau de Bure in France and the 12-meter telescope at Kitt Peak Observatory that are planned to join the EHT in 2020, and the Africa Millimetre Telescope (AMT), which is planned to be built on the Gamsberg in Namibia. Comparing this image to the simulated EHI images in Figure 7, it can be concluded that both the two- and three-satellite setups considered here offer the potential to reconstruct images that are significantly sharper and show less spurious substructure than images reconstructed from future ground-based observations. This enhancement is possible due to the higher frequency, longer baselines, and dense and isotropic $(u, v)$-coverage attainable for the EHI system. It should be noted that the EHT is able to generate images in less observing time because of the use of high-sensitivity stations like ALMA and the shorter $(u, v)$-plane filling timescale. The resolution may also be further improved with the planned upgrade to 345 GHz.

The images shown in this section were made using time-averaged source models. We refer to Ref. [46] for reconstructions from time-variable source models. After integrating and averaging for several spiral iterations with the two-satellite phase-stable configuration, the reconstructed images from a running GRMHD movie are of similar quality as for the time-averaged model.

### 4.2 Simulated Images of M87* for Case of Phase-stable Interferometer

EHI reconstructions of the time-averaged GRMHD model of M87* (Figure 2) are shown in Figure 10. The EHI observations were simulated using the same system parameters as for the two-satellite configuration observing Sgr A* (Figure 7, top row). The reconstructions show that the black hole shadow of M87* could be reconstructed with similar quality as that of Sgr A*.

## 5 Conclusions

In this paper, we have translated the science goals of imaging Sagittarius A* and M87* into technical requirements for a space-space VLBI system that observes in the 0.6 mm band with baseline lengths up to 25 000 km.

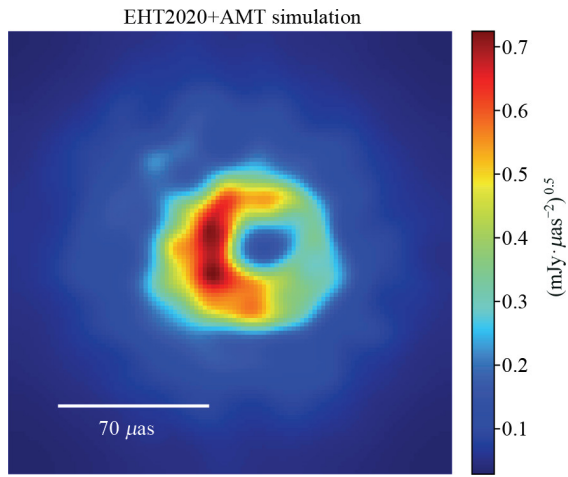

**Fig. 9**   Image reconstruction of the Sgr A* GRMHD model 39 at 230 GHz, from a simulated observation with the coverage of the EHT2020+AMT array. The observation was simulated with realistic observation and calibration effects using the SYMBA pipeline[95], and the image was reconstructed using the fiducial eht-imaging[93,94] script used in Ref. [9]

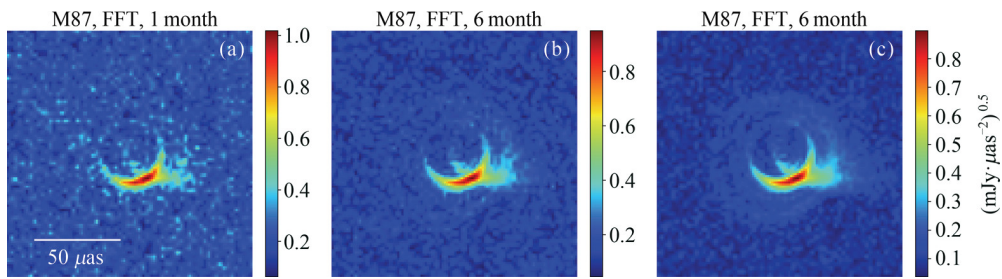

**Fig. 10**   EHI reconstructions of the time-averaged M87* model[40] at 557 GHz after 1, 6, and 24 months of integration in the phase-stable interferometer formed by the two-satellite configuration

We have considered a system architecture that allows all space components to be deployed using a single launch vehicle, in order to limit the mission budget cost efficient. The principal components of this 0.6 mm wavelength interferometer involve 4 m main reflectors, a system noise temperature of 150 K at the SSB receiver case, and a stability of ultra-stable oscillators below $5\times10^{-15}$ (Hz/Hz at 60 s). A sky bandwidth of 5 GHz is considered at 2-bits sampling.

The orbital plane orientation and the intended range of orbit semi-major axes are given. The chosen orbits and system architecture are both in line with the EHI concept, generating a dense and homogeneous filling of the $(u, v)$-plane out to the maximum baseline length. The synthesized antenna pattern is suitable for imaging without needing $(u, v)$-tapering or beam deconvolution.

Given the expected source geometry of the main targets, an effective integration time of multiple hours would be needed on the longest baselines in order to obtain successful fringe detections. The $(u, v)$-smearing prevents such a long integration time from being viable.

Independent LOs could be coherent beyond 60 s. Nevertheless, this coherence time is shorter than the $(u, v)$-cell crossing time. The three satellite configuration might allow the phase closure and the bispectrum could be accumulated in each $(u, v)$-cell over many realisations of the spiral. However, this configuration gives only relatively low angular resolution images because of the inefficiency of integrating bispectra with SNR $\ll 1$ at long baselines. To achieve the high resolution imaging required for the science goals, a phase stable interferometer is needed with coherence times longer than $(u, v)$-cell crossing time, approaching one spiral period. For this a shared LO scheme and excellent relative position determination are needed. At present, the ultimate capability of both a LO sharing and a-posteriori relative position knowledge are hard to estimate. It is therefore unclear without more work whether the extremely long stability timescales needed for (near) phase stable imaging with a 2 or 3 satellite system are feasible.

An avenue for measurement and estimation of the interferometer baseline state vector (and its derivatives) is proposed. The inter-satellite metrology itself is promising for delivering the range and the range-rate measurement accuracies on sub-mm scales, but the accuracy to which the full 3D baseline geometry can be measured is the subject of active study.

As the real-time knowledge on the baseline geometry will have a limited accuracy, a two-stage correlation approach is likely necessary. Such a two-stage correlation approach includes initial cross-correlation on-board across wide search windows in DnDR for the purpose of downlink data compression. The final selection of the DnDR solutions is done on-ground using the more accurate orbit propagation information that is available a-posteriori, so that multiple consecutive integration periods can be used to obtain a single fringe detection using modeled trends for DnDR.

In order to improve the SNR on long baselines further, multiple visibility measurements in the same $(u, v)$-grid cell can be averaged over longer time scales - either by averaging the visibilities directly or by using closure phases. We give simulated image reconstructions for different assumptions for both of these cases.

The paper provides the current progress on the critical elements of the Event Horizon Imager and indicates the foreseen avenue for further investigation on LO sharing and a two-stage correlation scheme.

**Acknowledgements** The authors are grateful to the following people. From JIVE ERIC and TU-Delft to Gurvits L. From ALMA to de Graauw T who suggested on metrology, candidate science goal, and move towards involving ground segment. From Institute of Space Studies of Catalonia to Li W who simulated the GNSS visibility at the initial configuration of 2 navigation antennas. From TESAT-Spacecom to Heine F who supported optical inter-satellite links. From ESA to Sodnik Z, Karafolas N, Murphy E, Haagmans R, Siemes C, Massotti L for support in optic communication and metrology incl. gravimetry; to Bras S for review and suggestion on the inertial navigation sensors; to Piironen P, Tuomas Kangas V, and Barre H who advised on observation front ends; to Ayllon N who suggested on microwave equipment; to de Maagt P, Paquay M, Salghetti Drioli L, Tauber J for their support on the parameters of antennas.

We acknowledge thoughtful suggestions of *Chinese Journal of Space Science*. We acknowledge EHT internal referee Johnson M.